\documentclass[3p,times,procedia]{elsarticle}
\usepackage{ecrc}

\volume{00}
\firstpage{1}
\journalname{Nuclear Physics A}
\runauth{U. Heinz et al.}
\jid{nupha}
\jnltitlelogo{Nuclear Physics A}

\usepackage{amssymb}
\usepackage{lineno}
\usepackage{graphicx}
\usepackage{bm}
\usepackage{amsmath,latexsym}
\usepackage[usenames]{color}
\usepackage{subfigure}
\usepackage{subfigure}
\usepackage{slashed}
\usepackage{multirow,array}
\usepackage{mathtools}
\usepackage{mathrsfs}
\usepackage[colorlinks=false,linktocpage=true]{hyperref}
\usepackage{hyperref}
\usepackage[utf8]{inputenc}
\usepackage{lipsum}
\usepackage[rightcaption]{sidecap}

\definecolor{darkblue}{RGB}{0,0,196}
\definecolor{darkred}{RGB}{196,0,0}

\biboptions{comma,sort&compress}
\usepackage[figuresright]{rotating}

\newcommand{\be}{\begin{equation}}
\newcommand{\ee}{\end{equation}}
\newcommand{\bea}{\begin{eqnarray}}
\newcommand{\eea}{\end{eqnarray}}
\newcommand{\beal}{\begin{align}}
\newcommand{\enal}{\end{align}}
\newcommand{\bs}{\begin{subequations}}
\newcommand{\es}{\end{subequations}}
\newcommand{\besp}{\begin{split}}
\newcommand{\eesp}{\end{split}}

\newcommand{\brs}{\beta_\mathrm{RS}}

\begin{document}

\begin{frontmatter}

\dochead{XXVIIth International Conference on Ultrarelativistic Nucleus-Nucleus Collisions\\ (Quark Matter 2018)}

\title{Thermalization \& hydrodynamics in Bjorken \& Gubser flows\tnoteref{tn0}}
\tnotetext[tn1]{Supported by DOE (award no.\,DE-SC0004286 and BEST Collaboration) and NSF (JETSCAPE Collaboration, ACI-1550223), as well as by the ExtreMe Matter Institute (EMMI) and the Fonds de Recherche du Qu\'ebec --- Nature et Technologies (FRQNT).}

\author[tata]{Chandrodoy Chattopadhyay}
\author[osu,emmi,cern]{Ulrich Heinz\footnote{Presenter. E-mail: heinz.9@osu.edu.}}
\author[tata]{Subrata Pal}
\author[osu]{Gojko Vujanovic}

\address[tata]{Department of Nuclear and Atomic Physics, Tata Institute of Fundamental Research, Homi Bhabha Road, Mumbai 400005, India}
\address[osu]{Department of Physics, The Ohio State University, Columbus, OH 43210, USA}
\address[emmi]{ExtreMe Matter Institute (EMMI), GSI Helmholtzzentrum f\"ur Schwerionenforschung, 
                          Planckstrasse 1, D-64291 Darmstadt, Germany}
\address[cern]{Theoretical Physics Department, CERN, CH-1211 Gen\`eve 23, Switzerland\\[-6ex]}

\begin{abstract}
The dynamical scaling behavior of hydrodynamic and non-hydrodynamic moments of the distribution function is studied using third-order Chapman-Enskog hydrodynamics and anisotropic hydrodynamics for systems undergoing Bjorken and Gubser expansions, where exact solutions of the Boltzmann equation in Relaxation Time Approximation (RTA) are available for comparison. While dimensionless quantities like normalized shear, pressure anisotropy and normalized entropy show at late times universal scaling relations with small (large) Knudsen number for Bjorken (Gubser) flows, dimensionful quantities like the entropy per unit rapidity do not. Although the two hydrodynamic approximation schemes describe the exact attractors for normalized shear with high accuracy, their description for the normalized entropy is less precise. We attribute this to non-negligible couplings to non-hydrodynamic modes in the entropy evolution.
\end{abstract}


\end{frontmatter}

\section{Introduction}
\label{sec:introduction}

Relativistic dissipative hydrodynamics has been remarkably successful in modeling the space-time evolution of the quark-gluon plasma formed in ultrarelativistic heavy ion collisions. However, the apparent success of hydrodynamics even in small systems \cite{Schenke:2017bog} and when space-time gradients are large has revived interest in testing the limits of applicability of the theory. Hydrodynamics is an effective theory of many-body systems in which microscopic interactions are averaged out, and its effective degrees of freedom are a small number of conserved charge currents coupled to dissipative fluxes. To gauge whether hydrodynamics adequately models the actual underlying dynamics one must compare the effective macroscopic description with a full solution of the microscopic dynamics. We here consider systems whose microscopic evolution can be described by relativistic kinetic theory. We study the evolution of systems described by the relativistic Boltzmann equation with a relaxation-type collision term 
(RTA), parametrized by a relaxation time $\tau_r=5(\eta/s)/T$ where $T$ is the temperature and $\eta/s$ the specific shear viscosity of the corresponding hydrodynamic fluid. For highly symmetric flow patterns introduced by Bjorken \cite{Bjorken:1982qr} and Gubser \cite{Gubser:2010ze} that share qualitative features with those encountered in heavy-ion collisions, this equation can be solved exactly \cite{Florkowski:2013lya, Denicol:2014xca}. 

As summarized in \cite{Heinz:2015gka} (to which we refer for additional references), different types of macroscopic hydrodynamic approximations of the Boltzmann equation can be obtained by splitting the single particle phase-space distribution function $f(x,p)$ into a leading-order contribution $f_0$ parametrized by hydrodynamic variables and a (small) deviation $\delta f$ describing (residual) dissipative corrections, and then expanding $\delta f$ either in gradients of the hydrodynamic parameters (Chapman-Enskog expansion, see e.g. \cite{Chapman_Cowling, Romatschke:2011qp, Jaiswal:2013vta, Chattopadhyay:2014lya} ) or in momentum moments (moment expansion, see e.g. \cite{Denicol:2012cn, Molnar:2016vvu}). The expansion is truncated at some order, and the Boltzmann equation is used for the truncated $\delta f$ to derive approximate relaxation-type equations of motion for the macroscopic dissipative flows, expressed as momentum moments of $\delta f$. After simplifying the resulting hydrodynamic equations for Bjorken and Gubser symmetries their solutions can the be compared with the hydrodynamic moments of the exact solution of the Boltzmann equation. Many such comparisons have been reported recently; we here focus on third-order Chapman-Enskog (CE) expansion \cite{Jaiswal:2013vta, Chattopadhyay:2014lya} and second-order anisotropic hydrodynamics (vaHydro) with $P_L$-matching \cite{Bazow:2013ifa, Tinti:2015xwa, Molnar:2016gwq, Martinez:2017ibh}. Additional results and references to earlier similar comparisons using different hydrodynamic approximations can be found in \cite{Heinz:2015gka, Chattopadhyay:2018apf}.

\section{Hydrodynamic evolution equations}
\label{sec2}

For massless systems, both Bjorken and Gubser symmetries restrict the dissipative flows to a single shear stress component. We use Milne coordinates $x^\mu=(\tau,x,y,\eta)$ for Bjorken flow \cite{Bjorken:1982qr} while for Gubser flow we work with de Sitter coordinates \cite{Gubser:2010ze} $\hat{x}^{\mu} = (\rho, \theta, \phi, \eta)$ on a curved spacetime, formed by Weyl rescaling of the flat Milne metric by $\tau^{-2}$. The coordinates are related by $\rho = - \sinh^{-1} \left[(1{-}q^2\tau^2{+}q^2 r^2)/(2q\tau) \right]$ and $\theta = \tan^{-1} \left[(2qr)/(1{+}q^2\tau^2{-}q^2 r^2) \right]$ where $1/q$ is an arbitrary length scale. Quantities in de Sitter coordinates are denoted by a hat and made unitless by scaling with appropriate powers of $\tau$. For the independent shear stress component we take $\pi \equiv -\tau^2\pi^{\eta\eta}$ in the Bjorken case and $\hat\pi\equiv\hat\pi^{\eta\eta}$ for Gubser flow.

{\bf 1. Bjorken flow:} A {\sl Chapman-Enskog expansion} around local thermal equilibrium up to third order in flow velocity gradients was performed for a conformal Boltzmann gas in \cite{Jaiswal:2013vta}. For Bjorken flow the corresponding equations for energy conservation and the relaxation of the shear stress reduce to \cite{Jaiswal:2013vta}
\begin{align}
\label{Bshear}
  \frac{d\epsilon}{d\tau} = -\frac{1}{\tau}\left(\frac{4}{3}\epsilon -\pi\right), \qquad
  \frac{d\pi}{d\tau} = - \frac{\pi}{\tau_\pi} 
   + \frac{1}{\tau}\left(\frac{4}{3}\beta_\pi - \lambda\pi - \chi\frac{\pi^2}{\beta_\pi}\right),
\end{align}
with transport coefficients $\tau_\pi=\tau_r,\ \lambda = 38/21,\ \chi = 72/245$, and $\beta_\pi = 4P/5$ where $P = \epsilon/3$ is the pressure. The local rest frame (LRF) entropy density has a non-equilibrium contribution, $s = s_\mathrm{eq} -\frac{3\beta}{8 \beta_\pi} \pi^2 - \frac{15\beta}{168\beta_\pi^2} \pi^3$ where $\beta=1/T$ is the inverse temperature \cite{Chattopadhyay:2014lya}. The LRF entropy current vanishes due to symmetry in Bjorken flow (but not in general \cite{Chattopadhyay:2014lya}).

In {\sl anisotropic hydrodynamics} one expands around an ellipsoidally deformed local momentum distribution, $f(x,p)\equiv f_a(x,p)+\delta\tilde{f}(x,p)$, where $f_a$ is the Romatschke-Strickland (RS) distribution \cite{Romatschke:2003ms}, $f_a = \exp\left[- \brs\sqrt{p_{\mu}p_{\nu}\Omega^{\mu\nu}}\right]$, with $\Omega^{\mu\nu}(x) = u^{\mu}(x)\,u^{\nu}(x) + \xi(x)\, l^{\mu}(x)\,l^{\mu}(x)$. The anisotropy parameter $\xi$ is obtained by Landau matching to the longitudinal pressure $P_L$ and, for Bjorken and Gubser symmetries, can be uniquely related to the shear stress $\pi$ resp. $\hat\pi$ \cite{Martinez:2017ibh}. For Bjorken flow one then finds instead of (\ref{Bshear}) the following modified evolution equation for the shear stress $\pi$:
\begin{align}
\label{Bshearaniso}
  \frac{d\pi}{d\tau} = - \frac{\pi}{\tau_\pi} 
   + \frac{1}{\tau}\left(\frac{25}{12}\beta_\pi - \lambda'\pi - I_{240}^\mathrm{RS}(\pi)\right),
\end{align}
where $\lambda'=8/3$ and $I_{240}^\mathrm{RS}$ is a thermodynamic integral over the RS distribution $f_a$ \cite{Molnar:2016gwq}, with $\xi$ expressed in terms of $\pi$ \cite{Martinez:2017ibh}. (The energy conservation law in (\ref{Bshear}) remains unchanged, of course.) The leading contribution $s_a$ to the entropy density is found to be $s_a(\tau) = 4/\left(\pi^2 \beta^3_\mathrm{RS}\sqrt{1{+}\xi}\right)$; the residual contribution to $s$ from $\delta \tilde f$ vanishes in the 14-moment approximation \cite{Chattopadhyay:2018apf}.

{\bf 2. Gubser flow:} For Gubser flow the energy conservation and shear relaxation equations from the {\sl third-order Chapman-Enskog expansion} reduce to \cite{Chattopadhyay:2018apf}
\begin{align}
\label{Gshear}
   \frac{d\hat{\epsilon}}{d\rho} = - \left( \frac{8}{3} \hat{\epsilon}  - \hat{\pi} \right) \tanh\rho, \qquad
   \frac{d\hat{\pi}}{d\rho} = - \frac{\hat{\pi}}{\hat{\tau}_\pi} 
   + \tanh\rho\left(\frac{4}{3} \hat{\beta}_\pi - \hat{\lambda} \hat{\pi} - \hat{\chi}\frac{\hat{\pi}^2}{\hat{\beta}_\pi}\right),
\end{align} 
with $\hat{\tau}_\pi{\,=\,}\hat{\tau}_r$, $\hat{\lambda}{\,=\,}46/21$, $\hat{\chi}{\,=\,}72/245$, and $\hat{\beta}_\pi{\,=\,}4\hat{P}/5$. The LRF entropy density is given by $\hat{s} = \hat{s}_\mathrm{eq} - \frac{3\,\hat{\beta}}{8\,\hat{\beta}_\pi} \hat{\pi}^2 + \frac{15\,\hat{\beta}}{168\,\hat{\beta}^2_\pi} \hat{\pi}^3$, and the LRF entropy flux again vanishes by symmetry. -- For {\sl anisotropic hydrodynamics} the energy conservation law is the same whereas the shear relaxation equation in (\ref{Gshear}) is modified to \cite{Martinez:2017ibh} 
\begin{align}
\label{AHshear}
   \frac{d \hat{\pi}}{d\rho} = -  \frac{\hat{\pi}}{\hat{\tau}_\pi} +
   \tanh\rho\left( \frac{4}{3} \hat{\beta}_\pi - \hat\lambda_a\hat\pi -\hat{I}_{240}^\mathrm{RS}(\hat\pi) \right).
\end{align}
The  integral $\hat{I}_{240}^\mathrm{RS}(\hat\pi)$ is defined in \cite{Martinez:2017ibh}. The expression for the entropy density is the same as for Bjorken flow above, with $\beta_\mathrm{RS}$ replaced by the rescaled $\hat\beta_\mathrm{RS}$.
 
\section{Results: Time evolutions of shear stress and entropy in Bjorken and Gubser flows}
\label{sec3}

\begin{SCfigure}[0.22][b!]
\includegraphics[width=0.79\linewidth]{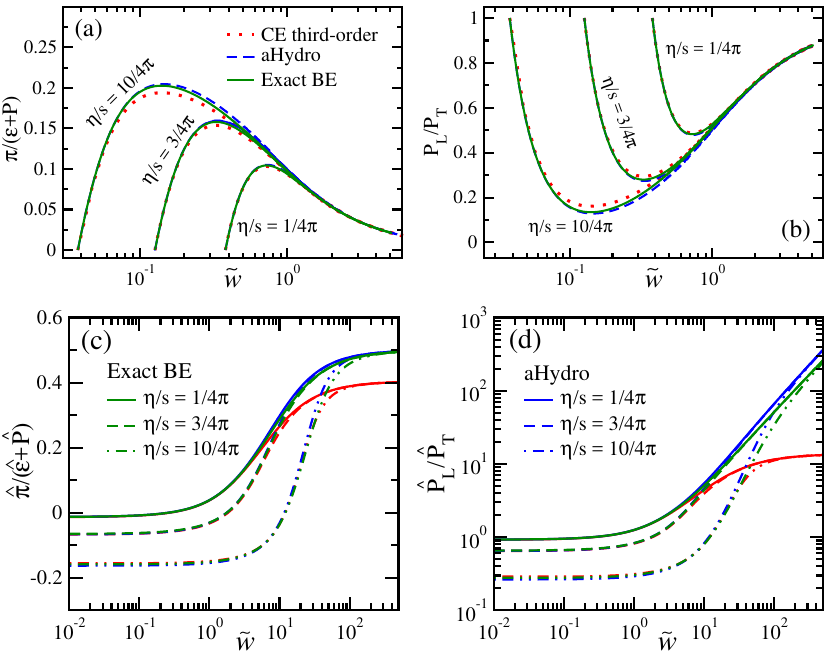}
\caption{Evolution with scaling variable $\tilde w$ (see text) of the normalized shear stress $\pi/(\epsilon{+}P)$ and the pressure anisotropy $P_L/P_T$ in Bjorken (panels a and b) and Gubser flows (panels c and d), for three values of the specific shear viscosity $4\pi\eta/s=1,\,3,$\, and 10. For Bjorken flow we used thermal equilibri\-um initial conditions of temperature $T_0{\,=\,}300$\,MeV at $\tau_0{\,=\,}0.25$\,fm/$c$. Gubser flow was initialized in thermal equilibrium with rescaled temperature $\hat T_0{\,=\,}0.002$ at $\rho_0{\,=\,}{-}10$. 
\vspace*{13mm}
\label{F1}
}
\end{SCfigure}

\begin{SCfigure}[0.21][t!]
\includegraphics[width=0.8\linewidth]{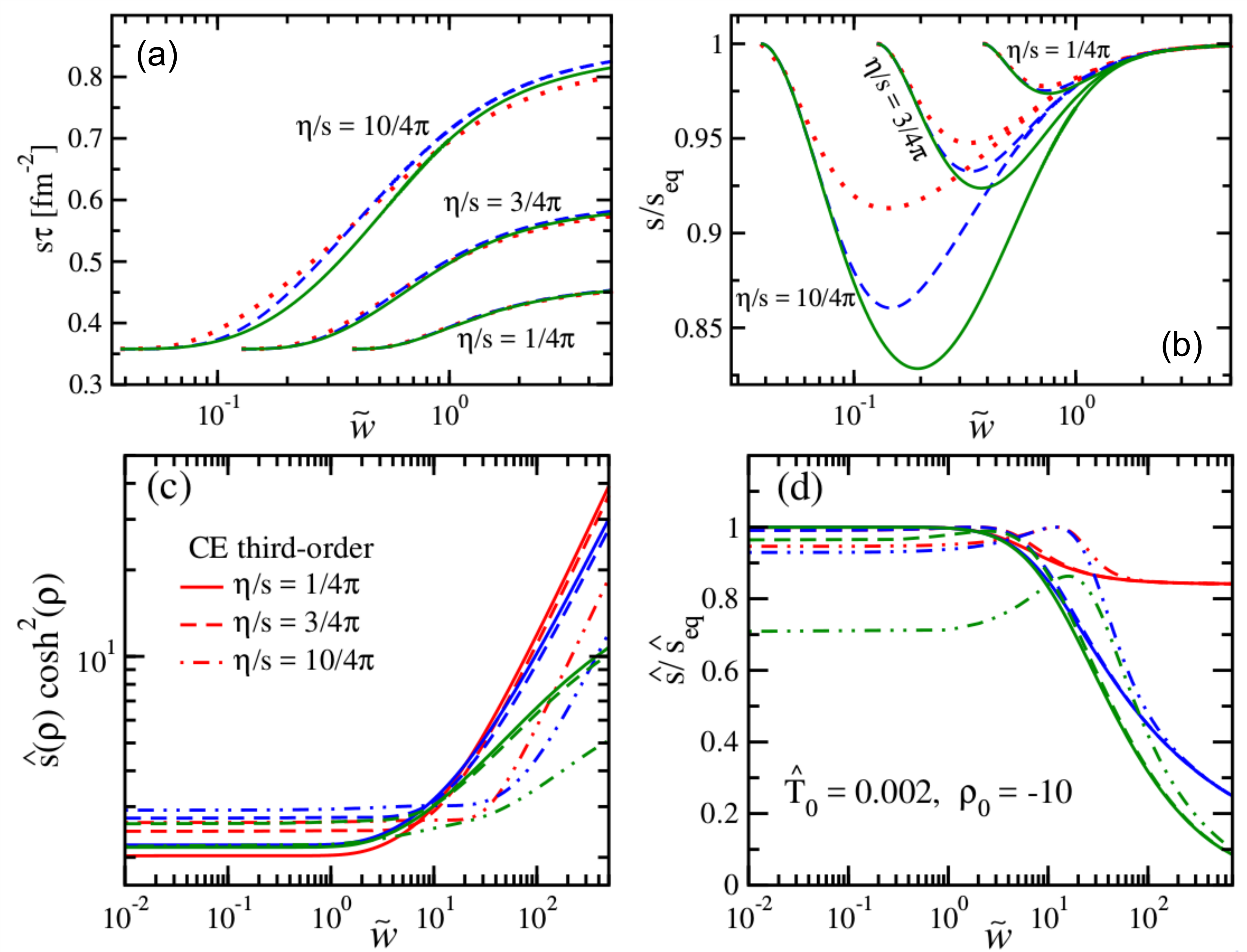}\vspace*{-4mm}\ 
\caption{Evolution with scaling variable $\tilde{w}$ (see text) of the entropy content and normalized entropy density in Bjorken (panels a and b) and Gubser flows (panels c and d). Same equilibrium initial conditions as in Fig.~\ref{F1}.
\vspace*{25mm}
\label{F2}
}
\end{SCfigure}

We now show comparisons of the time evolutions of the shear stress and entropy computed from the hydrodynamic equations in the preceding section with their exact evolutions from the exact solutions of the Boltzmann equation for Bjorken \cite{Florkowski:2013lya} and Gubser flows \cite{Denicol:2014xca} mentioned in the Introduction. In Figs.~\ref{F1}, \ref{F2}, the upper (lower) rows show results for Bjorken (Gubser) flow. We use rescaled time variables $\tilde w \equiv \tau T(\tau)/(4\pi\eta/s)=\mathrm{Kn}^{-1}$ (where Kn${\,=\,}\theta\,\tau_r$ is the Knudsen number) for Bjorken and $\tilde w\equiv (4\pi \eta/s) (2 \tanh \rho)/\hat{T}=\mathrm{Kn}$ for Gubser flow. For Bjorken flow the Knudsen number decreases at late times and the system approaches local thermal equilibrium: $\pi/(\epsilon{+}P)\to\bigl(\pi/(\epsilon{+}P)\bigr)_\mathrm{NS}=1/(3\pi\tilde w)$. For Gubser flow, the Knudsen number increases exponentially at late times, i.e. the system approaches free-streaming. Together these two systems allow to test the validity of hydrodynamic approximations in two diametrically opposite limits. The surprising observation from Fig.~\ref{F1} is that anisotropic hydrodynamics describe the exact evolution of the energy density \cite{Chattopadhyay:2018apf} and shear stress (or, equivalently, the pressure anisotropy $P_L/P_T$) extremely well {\em in both limits}, i.e. even at early times in Bjorken flow (when the Knudsen number Kn and the inverse Reynolds number Re$^{-1}{\,=\,}\pi/P$ are large) and at late times for Gubser flow (when the system approaches free-streaming and microscopic interactions cease). The 3rd-order Chapman-Enskog expansion performs almost equally well for Bjorken flow but for Gubser flow it fails to reproduce the correct asymptotic free-streaming value $\hat\pi/(\hat\epsilon{+}\hat{P})=0.5$ at large de Sitter times. Note that at large de Sitter times neither the exact solution nor the two hydrodynamic approximations follow the Navier-Stokes solution which diverges as $\bigl(\pi/(\epsilon{+}P)\bigr)_\mathrm{NS} = \tilde{w}/(6\pi)$.

In Fig.~\ref{F2} we study the evolution of local entropy content and normalised entropy density. For isentropic expansion the quantities $s\tau$ in Bjorken and $\hat{s} \cosh^2{\rho}$ in Gubser flow are constants of motion; in viscous evolution they increase. For Bjorken flow the entropy eventually saturates but for Gubser flow it continues to grow as the system moves further and further away from local equilibrium. At late times the normalised entropy density approaches a common attractor, just like the normalized shear stress and pressure anisotropy do in Fig.~\ref{F1}, albeit at somewhat larger values of the scaling variable $\tilde{w}$. However, for the entropy the late-time attractors of the hydrodynamic approximations differ from that of the exact solution, more so for 3rd-order CE than for anisotropic hydrodynamics.\\[-3ex]

\bibliographystyle{elsarticle-num}
\bibliography{Heinz_QM18}


\end{document}